# Variation of Chaotic Behaviour of Wind Speed Oscillations Across Indian Subcontinent

G. V. Drisya

Department of Futures Studies, University of Kerala, Kariavattom,
Trivandrum , Kerala - 695581, India
drisyavictoria@gmail.com

Alex Thumba

Thumba House, Piravom
Ernakulam, Kerala - 686664, India
alexthumba@gmail.com

**Abstract**
Modelling the intermittency of wind speed has got significant relevance on many fields, including wind power generation and distribution. Most of the available modelling techniques assume the temporal fluctuations in the wind is stochastic in nature and the underlying dynamics is best described by statistical methods or the probabilistic distribution. The advent of chaos theory have changed the perception about irregular fluctuations of dynamic systems and it has demonstrated that random-like fluctuations can also arise from deterministic chaotic systems. In this paper, we have analysed the deterministic nature of apparent random fluctuations seen in the daily average wind speed with the help of nonlinear time series analysis tools utilising wind speed data measured at nine typical locations over the Indian subcontinent from 2005 to 2015. The values of significant chaotic quantifiers obtained from the analysis clearly show the deterministic, low-dimensional and chaotic nature of wind speed dynamics in all these locations.

## 1. Introduction

Energy and economic growth are considered to be complementary to each other. While the energy contribution can stimulate economic growth in both direct and indirect ways by supporting the industrialisation phase of a region, economic development can engender increased demand for energy forcing the authorities to identify and develop new energy technologies. Sustainable energy resources including wind energy technologies are gaining much attention in recent days since the traditional non-renewable resources are depleting at a faster rate and very much hazardous to the environment. As the largest contributor and the fastest growing resource among renewable, wind energy is expected to continue its rapid growth for some decades and in worldwide the international wind community is monitoring advancements in any technology with an expectation of increasing the annual energy capture and driving down the cost of wind energy through R&D (IEA,2013). Characterization of the wind resource has already been identified by IEA as one of the strategic research areas by which remarkable cost reduction and optimal site assessment are possible. Short-term forecasting of wind is an important aspect of wind resource characterization, and an effective representation of the wind resource can improve the forecasting accuracy resulting in a more precise plant performance.

A number of modelling techniques can be found both in literature and practice in the aim of improved wind energy acquisition and utilisation (Kavasseri et al., 2009; Elliott, 2004; Finzi et al., 1984; Celik, 2004). Most of the tools reported for this purpose consider wind speed as a random process and used statistical methods like autoregressive models and probability distribution function to describe the underlying dynamics (Kamal et al., 1997; Kavasseri et al.,2009; Hennessey Jr, 1977; Cadenas et al., 2007; Celik, 2004; Mathew et al., 2011). In 1990, Elman introduced the concept of networks with memory which is capable of predicting the future behaviour from its previous responses. This concept is commonly known as artificial neural networks (ANN) and widely used for classification and prediction problems. ANN and its variants are widely used for making short-term wind speed and power predictions (Mohandes et al., 1998; Monfared et al., 2009; Mabel et al., 2009; Gomes et al., 2012; Beyer et al., 1994; Bechrakis et al., 1998). Support vector machines are supervised learning techniques for classification and regression and it identifies the best hyperplane with the maximum margin between the two classes (Steinwart et al., 2008). Use of SVM for predicting wind speed and power one step ahead has also been reported (Zhou et al., 2011; Zeng et al., 2011). Various studies of the effect of hybrid models, which combine different forecasting models for predicting the wind speed can also found in the literature (Soman et al., 2010; Liu et al., 2014; Haque et al., 2013). Despite all the efforts, none of the methods performed well enough in comparison with the persistence method which assumes the wind speed are persistent (Sfetsos, 2000). Undoubtedly highly intermittent nature of wind speed makes the system much difficult to model and exploring the sources of this random fluctuation can aid in better characterization and modelling of the wind resource.

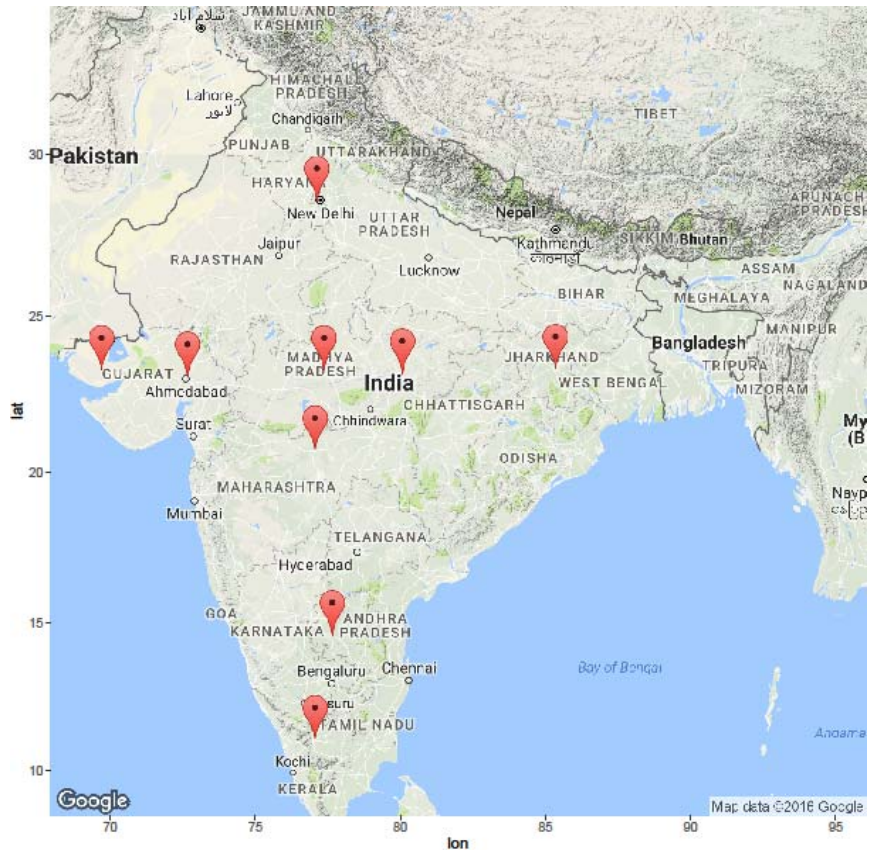

Figure 1: Locations at which daily mean wind speed (DMWS) data is measured.

Most of the reported wind speed modelling techniques assume the system as a stochastic process because of the presence of highly irregular fluctuations in the data. Since these fluctuations can also arise from a deterministically chaotic system, it is worth investigating whether the underlying dynamics is stochastic or deterministic. Palmer et al. 1995 analysed several time series of the horizontal wind speed and X-band Doppler radar signals measured concurrently over the ocean surface for nonlinearity and turned up with the result of the low-dimensional dynamical behaviour of both the systems. For a limited period, they could also obtain a higher prediction correlation coefficient with a neural network deterministic model. Ragwitz et al., (2007) attempted a comparison between the stochastic and deterministic model for wind speed time series. Even though Ragwitz reported no improvement in prediction accuracy by using nonlinear models, they observed sufficiently greater accuracy in predicting wind gusts. Hirata et al., (2008)proposed a prediction framework for wind direction based on a two-dimensional wind vector representation and they observed the time series data used for all these studies are too short. The analysis did by Martin et al., (1999) had used fairly large enough hourly data of wind speed time series measured over 7 years and expressed it as a sum of the deterministic component and a probabilistic or stochastic component. The analysis came up with the evidence of strong 1-year, 24-hour and 12-hour periodicity in deterministic components. These diurnal, yearly and semi-diurnal periodicities in wind time series are natural earth cycles and have been reported by Brett et al., (1991) and Gavaldà et al., (1992) and the presence of a periodic component in the data is a clear evidence of deterministic nature of the system. Nevertheless, the authors are vague in stating whether the ostensible random fluctuations are strictly from the stochastic process or arising out of chaotic underlying dynamics.

Apart from the local topography, earth rotation and solar heating are the major causes of surface wind blowing on earth and undoubtedly earth revolution is deterministic in nature. Many authors have argued solar radiation as a stochastic process and hence the wind can be modelled better with stochastic factors. However the recent studies on several other atmospheric parameters suggest the possibility of the other way round. Kumar et al., (2004) shown the strong chaotic nature of underlying dynamics of Total Electron Content (TEC), which is strongly influenced by the solar radiation. Assuming the surface wind as a similar parameter, Sreelekshmi et al. (2012) have done a preliminary analysis in this direction using 10 years daily mean wind speed (DMWS) measured at Thiruvananthapuram, Kerala, India. Their analysis of wind speed data in Thiruvananthapuram ($8.483^0$ N, $76.950^0$ E) using nonlinear time series analysis tools implemented in the TISEAN package (Hegger et al., 1999) reveals the possibility of the deterministic but the chaotic behaviour of the underlying dynamics of apparent random-like oscillations of wind speed measurements. Their assessments are based on a single location. As noted earlier, the wind speed dynamics is highly dependent on local topography. To make the affirmation further stronger, we have done a detailed analysis of DMWS data measured at various locations in India and the results are reported in this work. We have also examined the latitudinal and longitudinal variation in chaotic behaviour of the wind speed in terms of the measurements of some nonlinear quantifiers like lyapunov exponent, and correlation sum etc.. While the investigation of latitudinal variation is done by collecting DMWS data from five locations with a fixed latitude and varying longitude, longitudinal variation is done by collecting DMWS data from five locations with a fixed longitude and varying latitude. Figure 1. Shows the selected locations and the detailed information about these locations are given in Table 1.

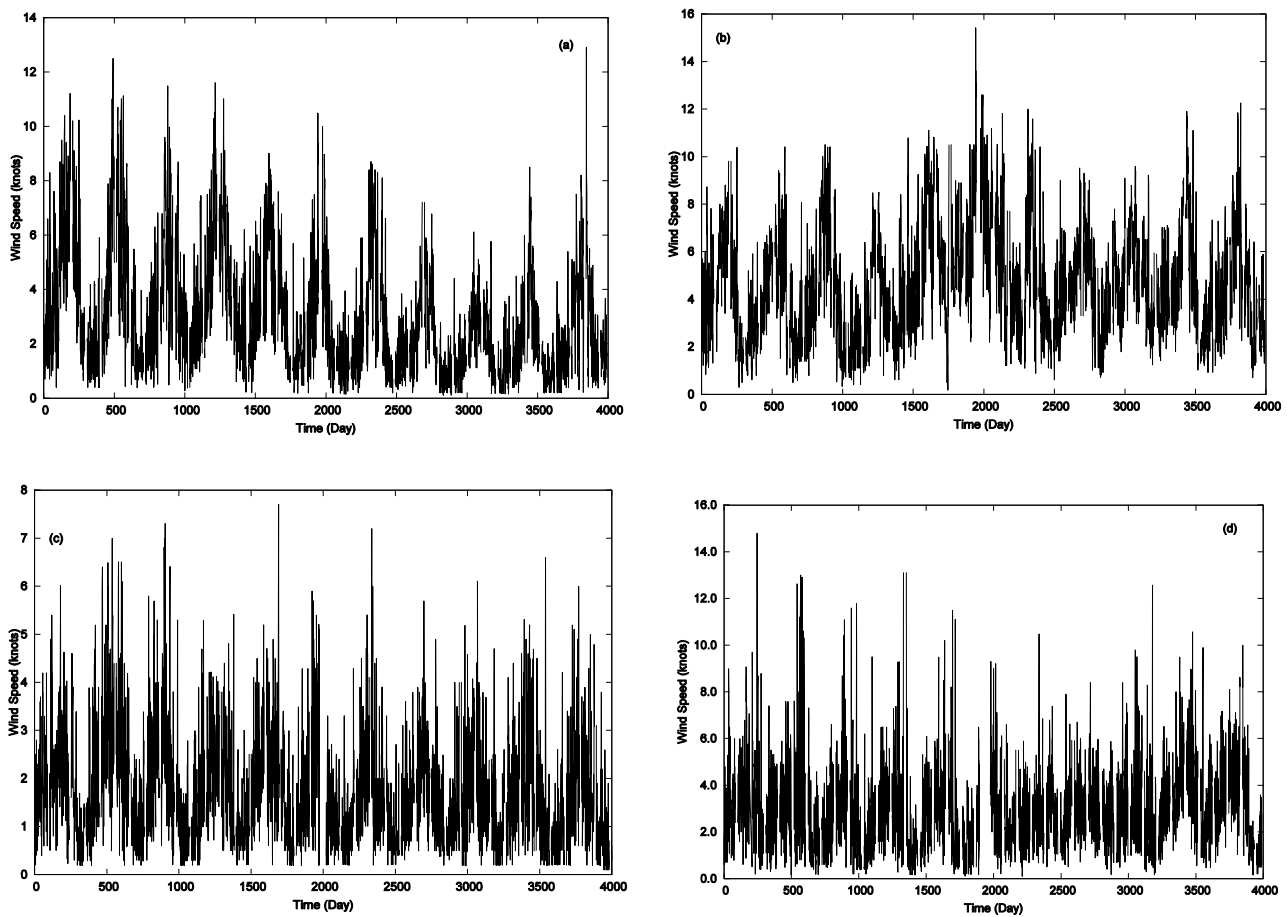

Figure 2: Time series of daily mean wind speed (DMWS) in knots measured across Indian subcontinent at locations (a) Bhuj (b) Ahemadabad (c) Jabalpur (d) Birsa Munda Airport.

## 2. Chaotic System

Daily mean wind speed measurements from different locations in India are plotted in Figure: 2. Random like temporal fluctuations are clearly evident in all the figures and as we discussed above these fluctuations can arise not only from a noisy stochastic system but also from a sensitive low-dimensional deterministic system.

Evidence of the chaotic behaviour of the dynamic system are first reported by Lorenz, (1963) and he well explained how a deterministic system is evolving into quite complex unpredictable states in the long term. Hence, before concluding fluctuations in the measured DMWS is purely random, a detailed analysis is necessary for exploring the sources of apparently random behaviour. Under some precise condition measures of complexity and predictability for characterising the system evolution are good tool to show whether the system follows a chaotic flow or not.

| Location | State | Latitude | Longitude |
|---|---|---|---|
| Bhuj | Gujarat | 23.287 | 69.67 |
| Ahemadabad | Gujarat | 23.077 | 72.634 |
| Bhopal | Madhya Pradesh | 23.287 | 77.337 |
| Jabalpur | Madhya Pradesh | 23.177 | 80.052 |
| Birsa Munda Airport | Jharkand | 23.314 | 85.321 |
| Coimbatore | Tamilnadu | 11.031 | 77.044 |
| Anantapur | Andhra Pradesh | 14.583 | 77.633 |
| Akola | Maharashtra | 20.7 | 77.033 |
| Indira Gandhi Airport | Delhi | 28.566 | 77.103 |

Table 1: Geographical information for the locations where wind speed data have been considered for analysis.

For any dynamical system time dependence of its states, which are represented by vector points in geometrical space, can be well described by the equation of motion and it is given by

$$\dot{x} \equiv \frac{dx}{dt} = f(x) \qquad (1)$$

where $x(t)$ is the state vector. For a dissipative system as $t \to \infty$ the trajectories that follow the system evolution will be attracted to a subset of phase space known as *attractor*. Characterisation of the attractor is an excellent technique to obtain the detailed information about properties of the system under consideration. Some dynamical systems are highly associated with chaotic behavior for its hypersensitivity to initial conditions. More precisely, trajectories following the states of the system that are originated from two points which were very close in phase space initially will deviate from each other in an exponential rate. In a longer period, even though the diverging trajectories may evolve separately without depending on each other and move forward in an uncorrelated manner, it will restrict themselves within the limits of a sub set of phase space. *Chaos is the bounded aperiodic behaviour in a deterministic system that shows sensitive dependence on initial conditions* (Alligood et al., 1997). In chaotic system, the adjacent trajectories following the system evolution may spread initially and eventually comes back to remain a bounded region. As $t \to \infty$ repeated spreading and folding of trajectories happens confining it in a specific region in phase space. This complex region, into which the states of the system is attracted as time evolve is known as attractor, and it maintains a definite geometry. Even though underlying dynamics of chaotic system are deterministic in nature, due to the property of sensitivity to initial condition, predictability is limited to short period. The restriction in long term prediction occurs due to the unavoidable measurement errors which will be amplified as time goes on and can cause exponential divergence of predicted trajectory from original one. Before the advent of chaos theory many chaotic systems producing apparent irregular behaviour were dubbed to be stochastic (Alligood et al., 1997; Ott, 2002).

## 3. Attractor Reconstruction

In many real world situations the dynamical system, as given in Eq. 1, or the state vector $x(t)$ may not be known or available, but only the measurements of a variable $y(t)$ equidistant in time *i.e.* a time series is accessible. The main objective in analysing such time series is to get insight into the underlying dynamical system. The first and foremost step in time series analysis is to reconstruct the dynamics of $x(t)$ on the attractor using the methodology known as attractor reconstruction, first suggested by Packard et al., (1980). The attractor generated by the $m$-dimensional delay vector constructed from y(t) at time interval (delay) $\tau$ is given by,

$$y(t) = (y(t), y(t + \tau), \ldots, y(t + (m-1)\tau)) \qquad (2)$$

and it will be topologically equivalent to the attractor of the state vector $x(t)$. The validity of the embedding $x(t) \to y(t)$ guaranteed by the embedding theorem of Takens, (1981) and its extensions by Sauer et al., (1991) and Sauer et al., (1993). For all values of delay $\tau$ and smooth measurement functions $h$, embedding is valid only if $m > 2D$, where $D$ is the box-counting dimension. Hence, the dynamical and geometrical characteristics of the original system $x(t)$ are preserved in the reconstructed space and can be computed from the measured time series $y(t)$ (Kantz et al., 2004; Ott et al., 1994). The attractor reconstruction depends on two parameters, the embedding dimension $m$ and delay $\tau$.

## 4. Nonlinear Time Series Analysis

A time series is the measurements of a time dependent variable at equal interval in time. Here we consider daily mean wind speed data, not the wind speeds measured at regular daily intervals. This averaging process contributes an additional additive noise apart from the measurement noise with zero mean and delta correlation. As a first step, we reduced the effect of the noise using the method proposed by Schreiber, (1993). Despite the apparent random-like fluctuations of the DMWS, time series plots in Figure. 2 shows strong annual variations. This was further confirmed by the space-time separation plot of each time series. A space-time separation plot depicts the relative separation in time of a pair points on a trajectory along the horizontal axis and their separation in space along the vertical axis. A space-time separation plot is useful in identifying the temporal correlations in a time series (Provenzale et al., 1992). Typical space-time separation plots are given in Figure. 3. In an epoch analysis, on each time series the modulation effect annual variations was reduced by deducting from each of the data points which are 365 days apart their average value (Kumar et al., 2004). The variation of 28 days arising from lunar influence was evident from the resulting time series and hence the procedure was repeated for each time series to reduce the effect of 28-day variations. Further analysis was carried out on each of the resultant denoised and detrended time series retaining temporal variations. The plots of the de-noised and de-trended time series of eight locations are given in Figure. 4 and their space-time separation plots are given in Figure. 5.

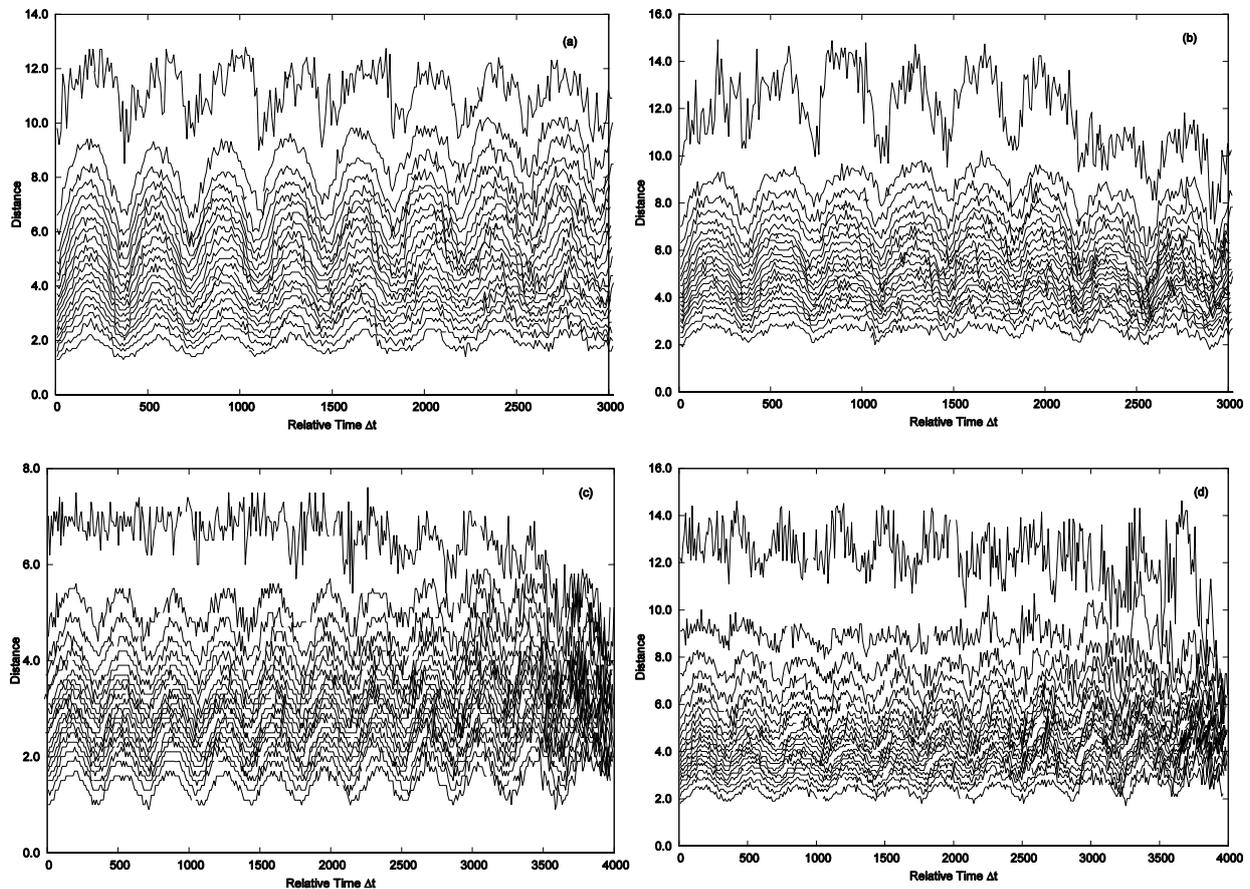

Figure 3: Space-time separation plot for the time series measured at locations (a) Bhuj (b) Ahemadabad (c) Jabalpur (d) Birsa Munda Airport.

The first step in the analysis of the denoised data is the reconstruction of the attractor as per the method discussed in the previous section for which one needs to estimate the suitable embedding parameters - delay $\tau$ and embedding dimension $m$. The choice of $\tau$ and $m$ crucial in inferring the result of the experimental measurements. Selection of small values of delay shall lead to highly correlated vectors $y(t)$ leading to unduly larger values for the correlation dimension and the choice of large values for delay can also lead to fairly uncorrelated components resulting in data randomly distributed in the embedding space (Kantz et al., 2004). Proper choice of the time delay is, therefore, important and a first guess of a suitable delay may be obtained from the autocorrelation function of the sample data $y_i$ given by

$$\rho(t) = \sum_i \frac{(y_i - \bar{y})(y_{i+\tau} - \bar{y})}{\sum_i (y_i - \bar{y})} \qquad (3)$$

where $\bar{y}$ is the sample mean (Kantz et al., 2004). A better method to fix the value the delay $\tau$ is to calculate the time-delayed mutual information suggested Fraser et al., (1986). This method also takes into account of the non-linear correlations.

In this method, a quantity called average mutual information is computed for various delays as a measure of the predictability of $y(t + \tau)$ given $y(\tau)$.

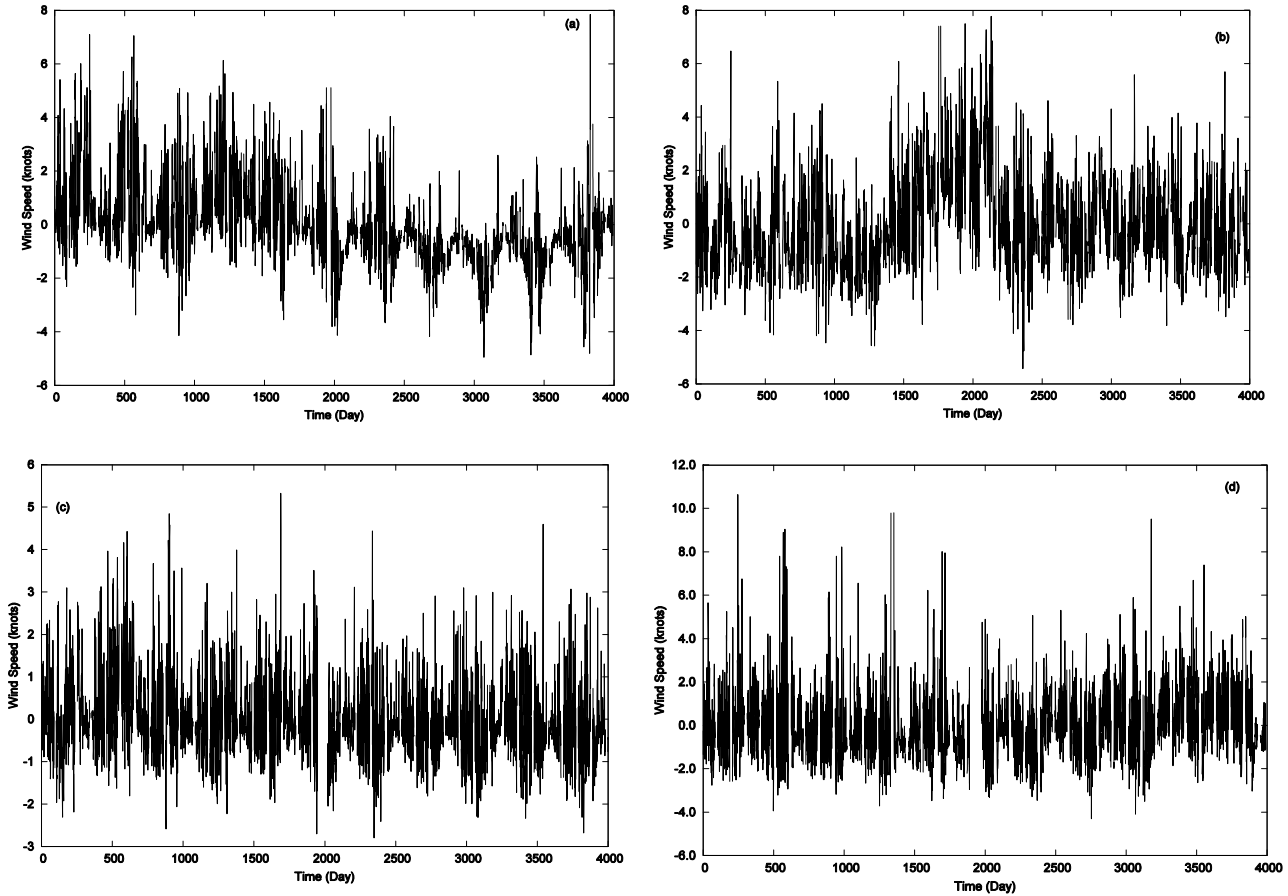

Figure 4: Noise reduced de-trended time series data measured at locations (a) Bhuj (b) Ahemadabad (c) Jabalpur (d) Birsa Munda Airport.

For a given delay, the mutual information $I(\tau)$ or a given delay $\tau$ is calculated by treating the sequences $y_i$ and $y_{i+\tau}$ as values of random variables $X$ and $Y$ with the formula,

$$I(\tau) = \sum_{y \in Y} \sum_{x \in X} p(x, y) \log_2 \left( \frac{p(x, y)}{p(x)p(y)} \right) \qquad (4)$$

Here, $p(x, y)$ is the joint probability mass function of X and Y with marginal p(x) and p(y). The probabilities can be estimated by constructing a histogram of the data points. The average mutual information of a time series $y(t)$ is computed for various values of the delay $\tau$ as a measure of the predictability of $y(t + \tau)$ given $y(\tau)$. The value of the average mutual information when plot against increasing values of delay $\tau$ exhibits a marked minimum and it can be good estimate of the optimal value of the delay parameter $\tau$. The average mutual information DMWS data indicates $\tau = 1$ can be a good choice for all the locations under study. Typical plots of the average mutual information for four different locations are given in Figure. 6. The plots of other locations also show similar features. We would like to emphasise that the product $m\tau$ is more important than values of $m$ and $\tau$ independently. The precise knowledge of $m$ is only required to estimate the dynamics with minimal computational effort (Kantz et al., 2004).

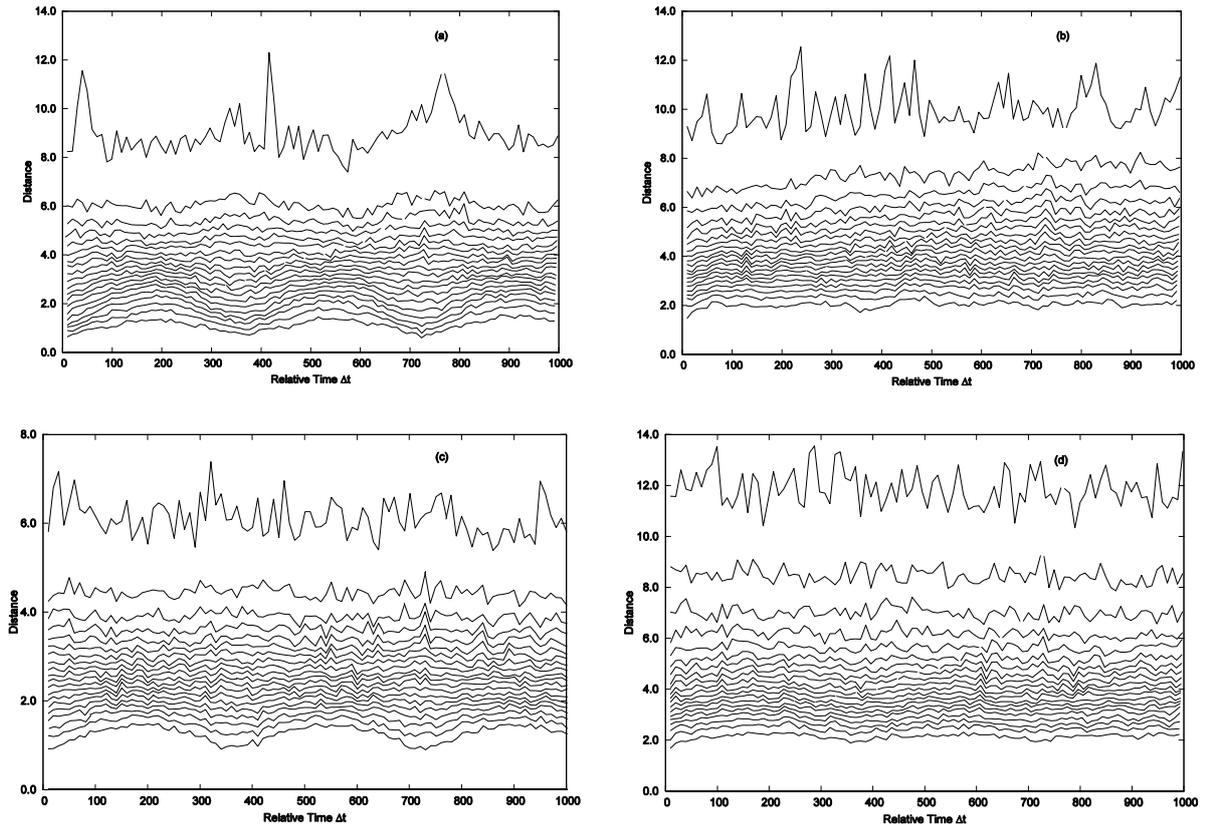

Figure 5: Space-time separation plot for the noise reduced de-trended time series data from locations (a) Bhuj (b) Ahemadabad (c) Jabalpur (d) Birsa Munda Airport.

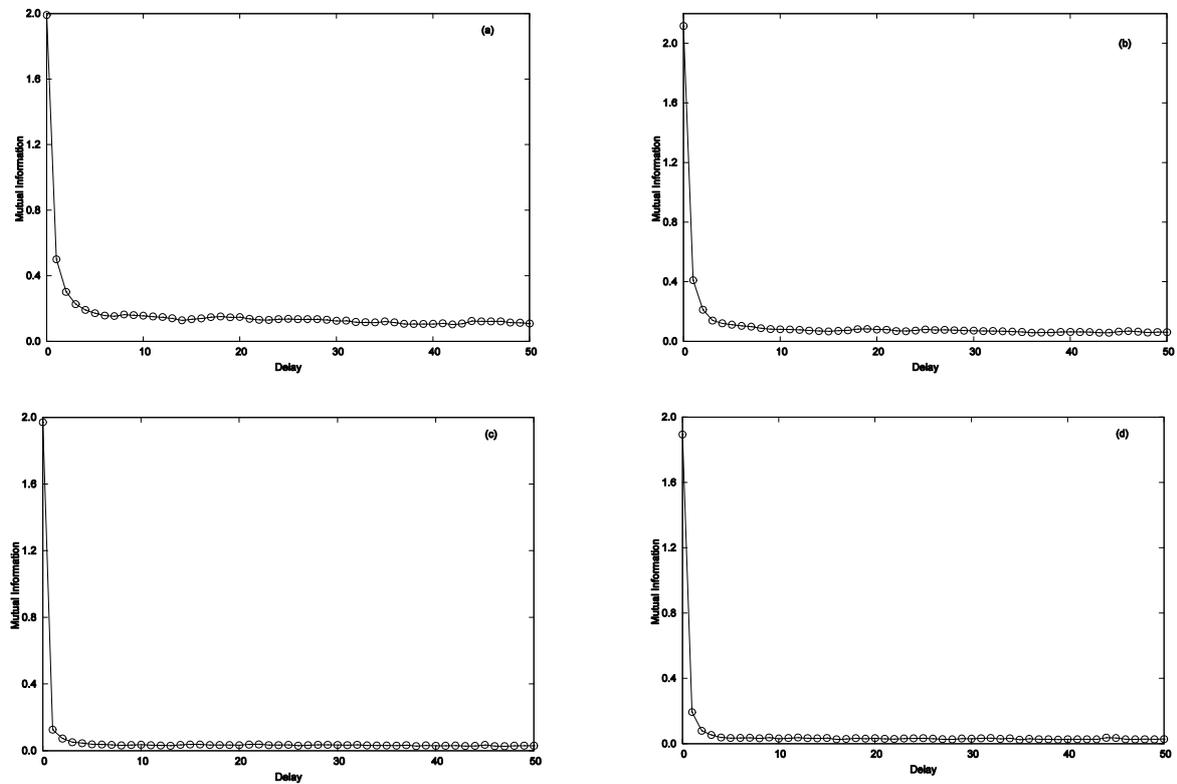

Figure 6: Mutual Information of the noise reduced de-trended time series data from locations (a) Bhuj (b) Ahemadabad (c) Jabalpur (d) Birsa Munda Airport.

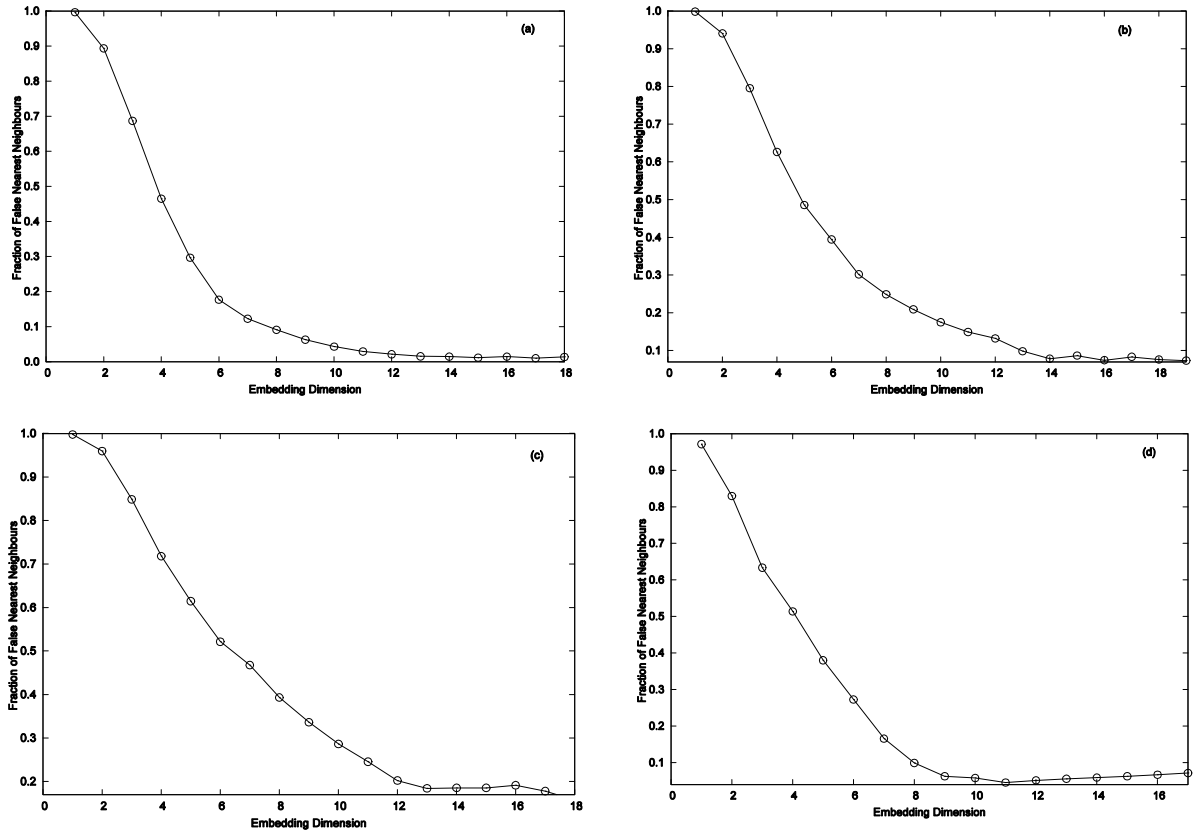

Figure 7: False nearest neighbours of the noise reduced de-trended time series (a) Bhuj (b) Ahemadabad (c) Jabalpur (d) Birsa Munda Airport.

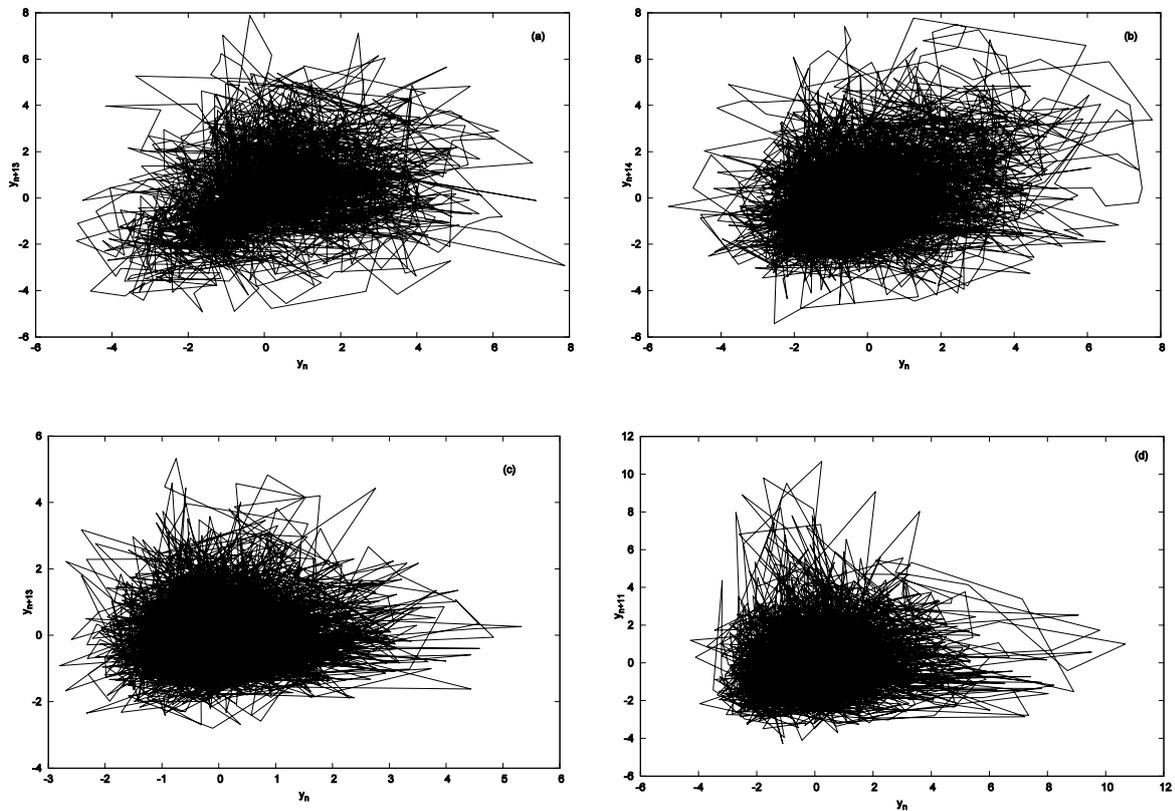

Figure 8: The delay representation of the denoised de-trended time series (a) Bhuj (b) Ahemadabad (c) Jabalpur (d) Birsa Munda Airport.

According to Kantz et al., (2004) the value of embedding dimension $m$, it should be large enough for the attractor to fully unfold in the embedding space. However, a value that is too large may cause the various algorithms to underperform (Kantz et al., 2004). Kannel et al. (1992) proposed a practical method to choose the right value of the embedding dimension $m$ by calculating the fraction of false neighbours as a function of $m$. False neighbours arise due to the crossing of trajectories when the attractor can not unfold its true geometry when the value of $m$ is not large enough as a result of projection onto a small dimensional space. To estimate a suitable value for the $m$ fraction of false neighbours are computed in progressively higher dimension until the difference becomes negligible. The first value of $m$ corresponding to the first minimum of the fraction of false neighbours indicates a suitable value for the embedding dimension. We have computed the fraction of false neighbours of DMWS data all locations considered, and plots of four of them are given in Figure. 7. It can be seen that values of $m > 13$ (in some case $m > 12$) the fraction of false neighbours become extremely small. The value $m = 14$ can be selected safely in all these locations. This shows preliminary indication that underlying dynamics is low dimensional in character across the locations.

The delay representation of the time series in four locations with $m = 14$ and $\tau = 1$ are given in Figure. 8. The definite structure in these figures is an indication of the deterministic nature of the underlying dynamics. The similar pattern is observed in other locations as well. A typical feature of a chaotic attractor is its self-similarity. Various dimension estimates, such as the box-counting dimension, the Hausdorff dimension etc., were introduced in the literature to quantify the structure of a chaotic attractor. All these quantities of dimension are the generalisation of the Euclidean definition of dimension to account the self-similar character of chaotic attractors. The dimension estimate of a chaotic attractor need not be an integer. Among all these most popular in literature is the correlation dimension introduced by Grassberger et al., (1983). The correlation integral $C(\epsilon)$, is the probability that a pair of points chosen randomly on the attractor is separated by a distance less than $\epsilon$. It is found that $C(\epsilon) \propto \epsilon^{D2}$ as $\epsilon \to 0$. Therefore, the correlation dimension can be computed from the slope of the curve of $\ln C(\epsilon)$ versus $\ln(\epsilon)$ given by

$$D_2 = \lim_{\epsilon \to 0} \frac{d \ln C(\epsilon)}{d \ln \epsilon} \tag{5}$$

For a single time series and $N$ data points of m-dimensional delay vectors $y_i$, the correlation integral $C(\epsilon)$ is approximated by the correlation sum $C(\epsilon, m)$ given by Kantz et al., (2004).

$$C(\epsilon, m) = \frac{2}{N(N-1)} \sum_{i=1}^{N} \sum_{j=i+1}^{N} \Theta(\epsilon - \|y_i - y_j\|) \tag{6}$$

for sufficiently large $N$, where $\Theta(a) = 1$ if $a > 0$, $\Theta(a) = 0$ if $a \leq 0$. In practice the local slopes

$$D_2(\epsilon, m) = \frac{d \ln C(\epsilon, m)}{d \ln \epsilon} \tag{7}$$

are computed and plots them as a function of $\epsilon$ for various $m$; the value corresponding to a plateau in the curves is estimated and identified as an approximate value of $D_2$.

However, only the spatial closeness of points should be accounted for in Eq. 7 whereas in some cases, it can be affected by the temporal closeness of points as well. To avoid this points which are closer in time by less than a *Theiler window $\omega$*, which is approximately equal to the product of the time delay and the embedding dimension, should be excluded from the computation of the correlation sum (Theiler, 1986). According to Hegger et al., (1999) the value of $\omega$ should be chosen generously.

The typical plots of the local slopes $D_2(\epsilon, m)$ is given in Figure. 9. We have calculated the correlation dimension of the DMWS data at all locations and is given in Table 2. The wind dynamics may be affected myriads of factors, but the estimated values of the correlation dimension show that the eventual behaviour characterised by the attractor is low dimensional.

The most striking feature of a chaotic system is its sensitivity to initial conditions. Therefore the trajectories starting from neighbouring initial conditions diverge exponentially as time passes. The average rate of divergence of nearby trajectories is quantified by what known as Lyapunov exponent. Positive Lyapunov exponent is a striking evidence for chaotic behaviour of system (Ott, 2002).

The growth of the divergence $n\delta(t)$ between two neighbouring trajectories is quantified by the maximum Lyapunov exponent $\lambda$, so that $\|\delta(t)\| = \|\delta(0)\|e^{\lambda t}$ and hence

$$\lambda = \lim_{t \to \infty} \frac{1}{t} \ln \frac{\|\delta(t)\|}{\|\delta(0)\|} \tag{8}$$

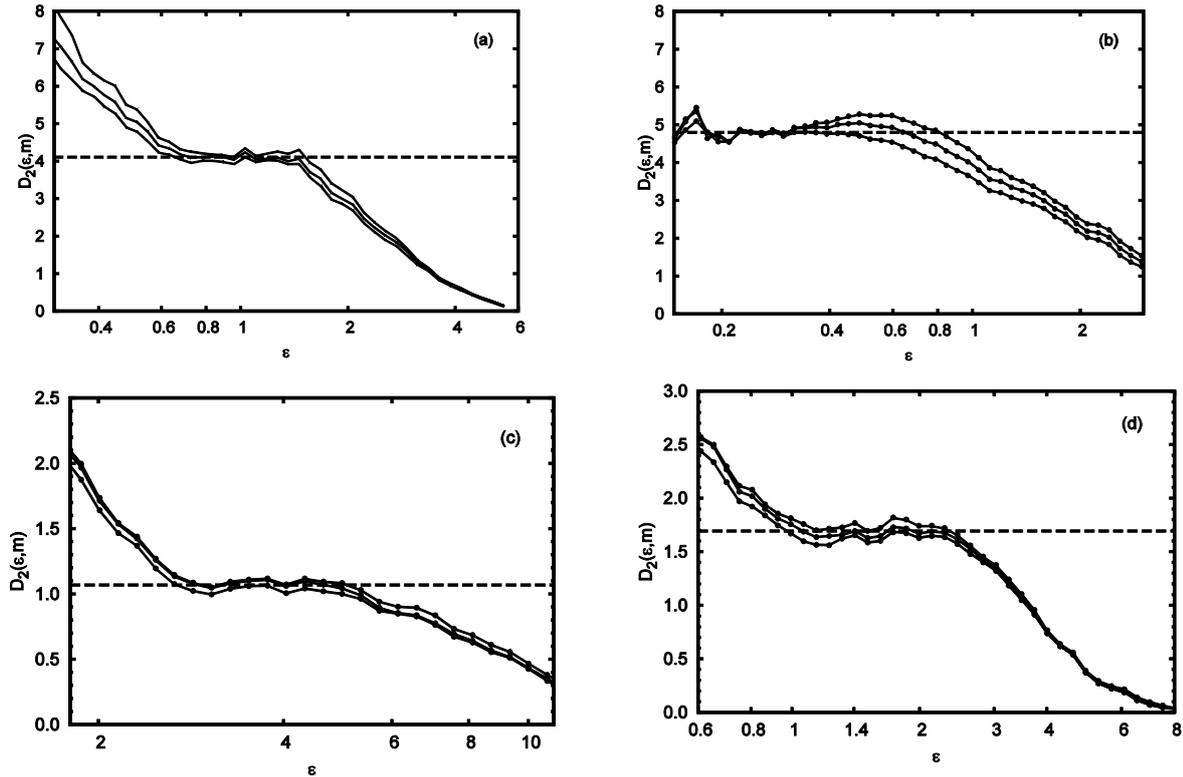

Figure 9: The local slopes $D_2(\epsilon, m)$ for the de-noised and de-trended time series of DMWS at (a) Jabalpur (b) Coimbatore (c) Anantapur and (d) Akola. The values of the correlation dimension are given in Table 2.

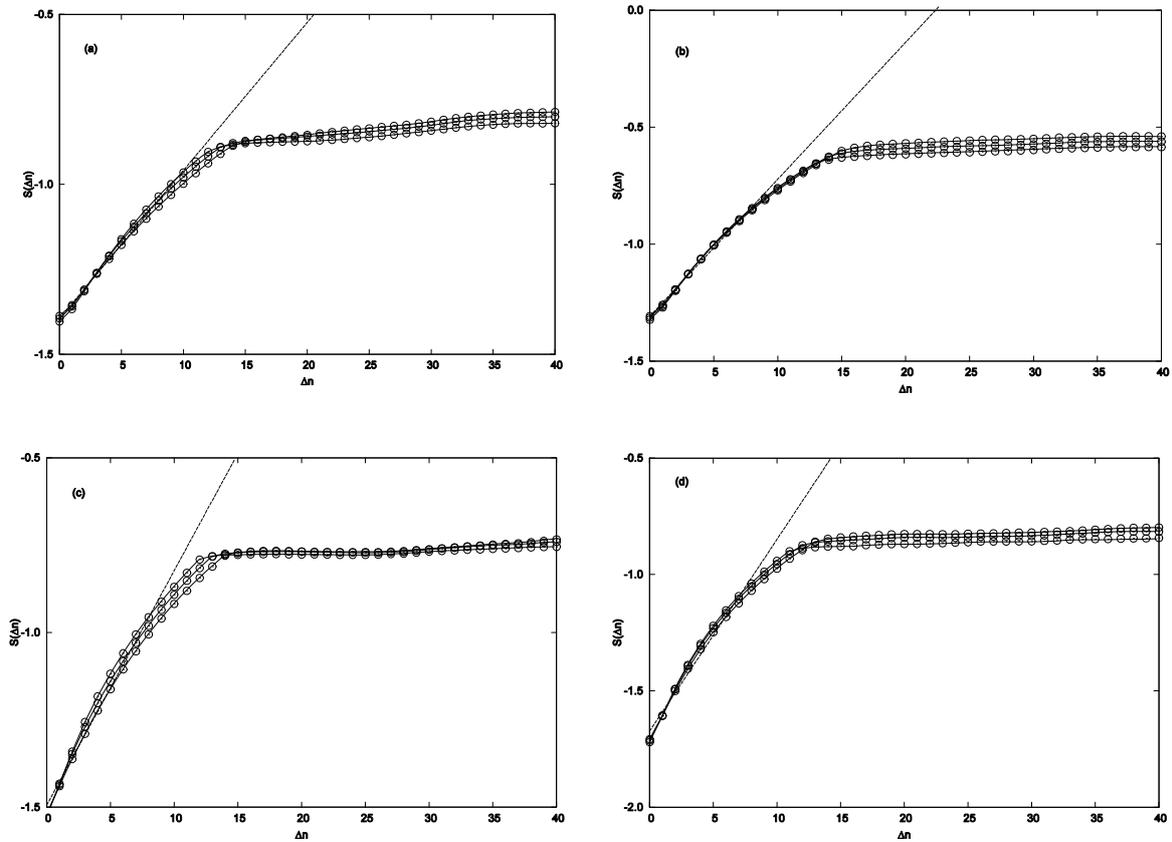

Figure 10: The curve of $S(\delta n)$ for various embedding dimension $m$. The maximum Lyapunov exponent of de-trended time series is the slope of the dashed line. (a) Bhuj (b) Ahemadabad (c) Jabalpur (d) Birsa Munda Airport.

| Location | Latitude (degree) | Longitude (degree) | Correlation Dimension | Lyapunov exponent |
|---|---|---|---|---|
| Bhuj | 23.287 | 69.67 | 2.2488 ± 0.0092 | 0.0434 ±0.0007 |
| Ahemadabad | 23.077 | 72.634 | 5.1174 ± 0.0587 | 0.0471 ±0.0013 |
| Bhopal | 23.267 | 77.337 | 2.8775 ± 0.0579 | 0.0642 ±0.0016 |
| Jabalpur | 23.177 | 80.052 | 4.1087 ±0.0265 | 0.0675 ±0.0018 |
| Birsa Munda Airport | 23.314 | 85.321 | 4.7908 ± 0.0331 | 0.0826 ±0.0022 |
| Coimbatore | 11.031 | 77.044 | 4.8001 ± 0.0213 | 0.0984 ±0.0013 |
| Anantapur | 14.583 | 77.633 | 1.0685 ± 0.0071 | 0.0778 ±0.0010 |
| Akola | 20.700 | 77.033 | 1.6936 ± 0.0116 | 0.0889 ±0.0011 |
| Indira Gandhi Airport | 28.566 | 77.103 | 5.9420 ± 0.1863 | 0.0742 ±0.0015 |

Table 2: The estimated values of the correlation dimension at various locations. The values appear to depend more on the local topography rather than geographical location.

The maximum Lyapunov exponent $\lambda$ can be estimated by plotting $\ln \delta(t)$ versus $t$, which should fall nearly in a straight line, the slope of which then gives an estimate of $\lambda$. Lyapunov exponent have preserved delay reconstruction as they are invariant under smooth transformations of the attractor, and hence they can be estimated from a time series. There are several algorithms for estimating the maximal Lyapunov exponent from time series. Kantz algorithm (Kantz, 1994; Kantz et al., 2004) is popular in the literature which starts with computing the sum

$$S(\Delta n) = \frac{1}{N} \sum_{n_0=1}^{N} \ln \left( \frac{1}{\|U(y_{n_0})\|} \sum_{y_n \in U(y_{n_0})} \|y_{n_0+\Delta n} - y_{n+\Delta n}\| \right) \quad (9)$$

for a point $y_{n_0}$ of the time series in the embedded space and over a neighbourhood $U(y_{n_0})$ of $y_{n_0}$ with diameter $\epsilon$. If the plot of $S(\Delta n)$ versus $\Delta n$ is linear over small $\Delta n$, for a reasonable range of $\epsilon$, and all have identical slope for sufficiently large values of the embedding dimension $m$, then that slope can be taken as an estimate of the maximum Lyapunov exponent (Kantz, 1994; Kantz et al., 2004).

The curves of $S(\Delta n)$ for $m = 14,15,16$ for various locations are given in Figure. 10. The estimated values of the maximum Lyapunov exponent are given in Table 2. The positive values of the maximum Lyapunov exponents show the underlying dynamics of wind speed variations in all these locations are chaotic in nature.

Many characteristics of a chaotic system are also mimicked by a color noise time series. Pavlos et al. (1992) have noted that phase randomization of a chaotic signal can destroy its profile whereas colour noise time series preserve its profile. We have compared the local slopes of logarithm $D_2(\epsilon, m)$ of the correlation sum original time series of each location and compared with its phase randomised time series. The phase randomised time series is obtained by obtaining the Fourier series representation original and reconstructing the time series after adding random phase distribution. Essentially we obtained plots like in Fig.11 in (Sreelekshmi et al., 2012). In each location, the chaotic profile of the time series was destroyed by phase randomization indicating the deterministic nature the given time series.

## 5. Surrogate Data Test

Stochastic systems driven by a linear Gaussian process distorted by some nonlinear process might also exhibit many features of a chaotic system. The main objective of surrogate data test is to ascertain that the complex behaviour exhibit by a time series is not stochastic. We validate further that, the results reported in the previous section were not arisen from a linear stochastic process by carrying out surrogate data test on DMWS data at all locations considered.

The method of surrogate data has been widely used for discriminating if the source of apparent random fluctuations in a time series is deterministic or stochastic (Theiler et al., 1992). It is a statistical test to formally reject the hypothesis that the observed time series arose from a linear noise process. The null hypothesis is first formulated that the observed time series is a random process and then an ensemble of random numbers, called surrogate data, consistent with the null hypothesis and otherwise similar to the original data were generated. Then one proceeds to test if a discriminating statistic such as correlation dimension or Lyapunov exponent computed from the surrogate data is significantly different from that of the original data. The null hypothesis is rejected if they significantly different.

For each time series of the DMWS measurements, we generated 40 surrogates by the Amplitude Adjusted Fourier Transform method proposed by Schreiber and Schmitz [51]. The surrogates preserve the amplitude distribution, power spectrum and autocorrelation of the original DMWS data, so that they can be treated as what the realisations satisfying the null hypothesis. To test null hypothesis we have employed both geometrical and dynamical characteristics such as the fraction of false nearest neighbours, the local slopes of the correlation sums and the curves of $S(\Delta n)$.

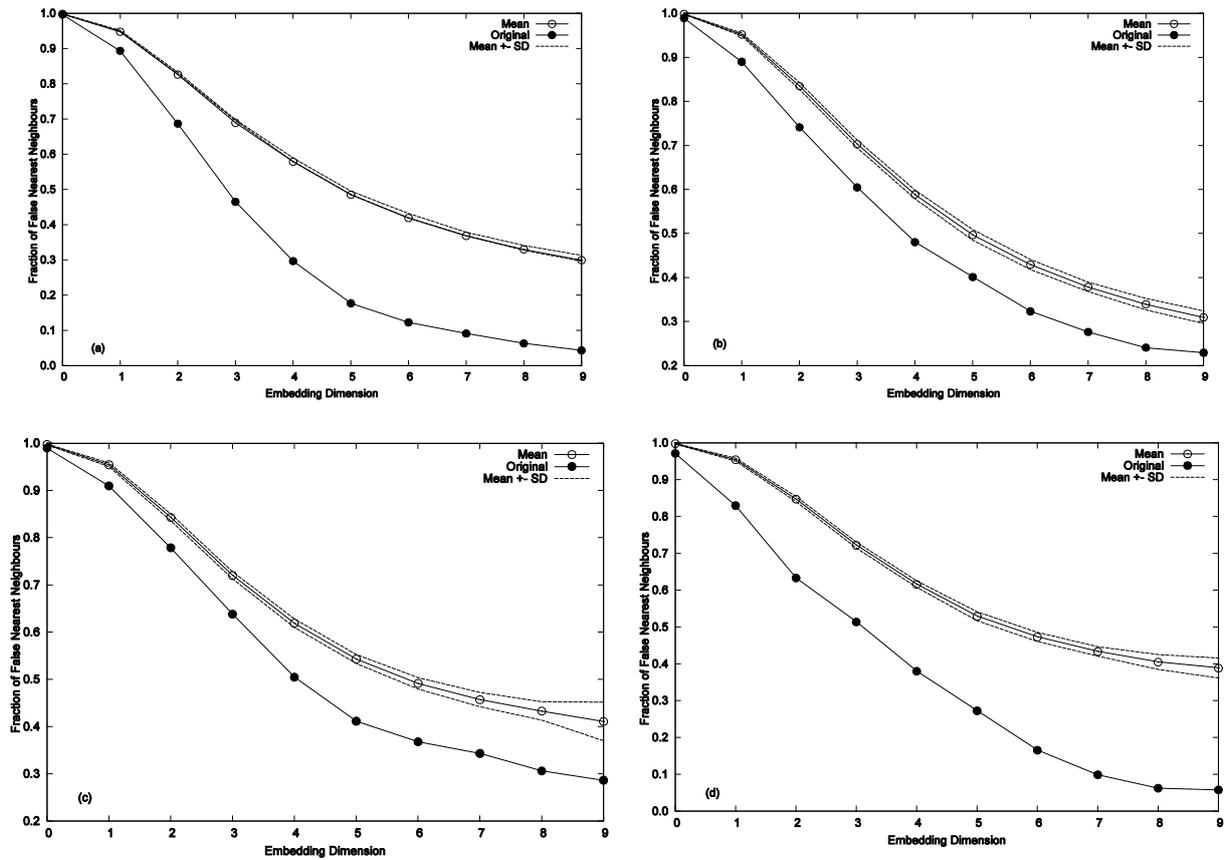

Figure 11: The mean values of the fraction of false nearest neighbours of the surrogates with standard deviation. (a) Bhuj (b) Ahemadabad (c) Jabalpur (d) Birsa Munda Airport.

The above characteristics are calculated for the data, both original and the surrogate, and the null hypothesis is accepted or rejected depending on the value of the significance of the difference given by Mitschke et al., (1993) and Pavlos et al., (1999)

$$S = \frac{\mu - \mu_{org}}{\sigma} \tag{10}$$

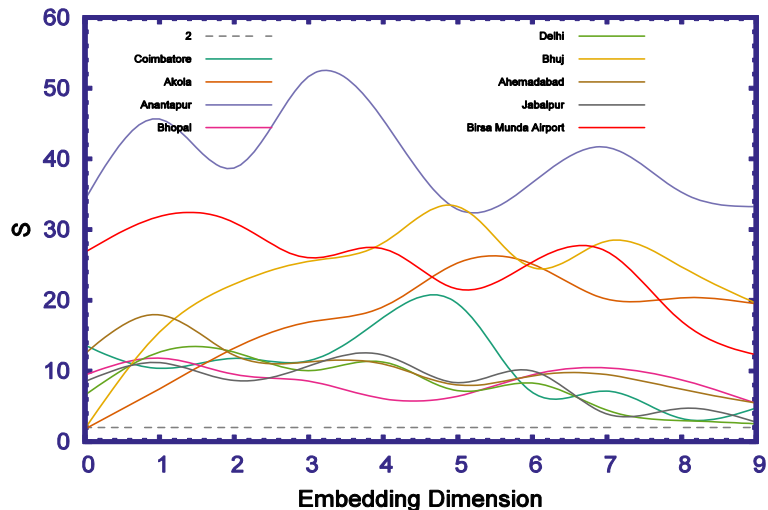

Figure 12: Plot of the significance of difference $S$ versus $m$ for different locations.

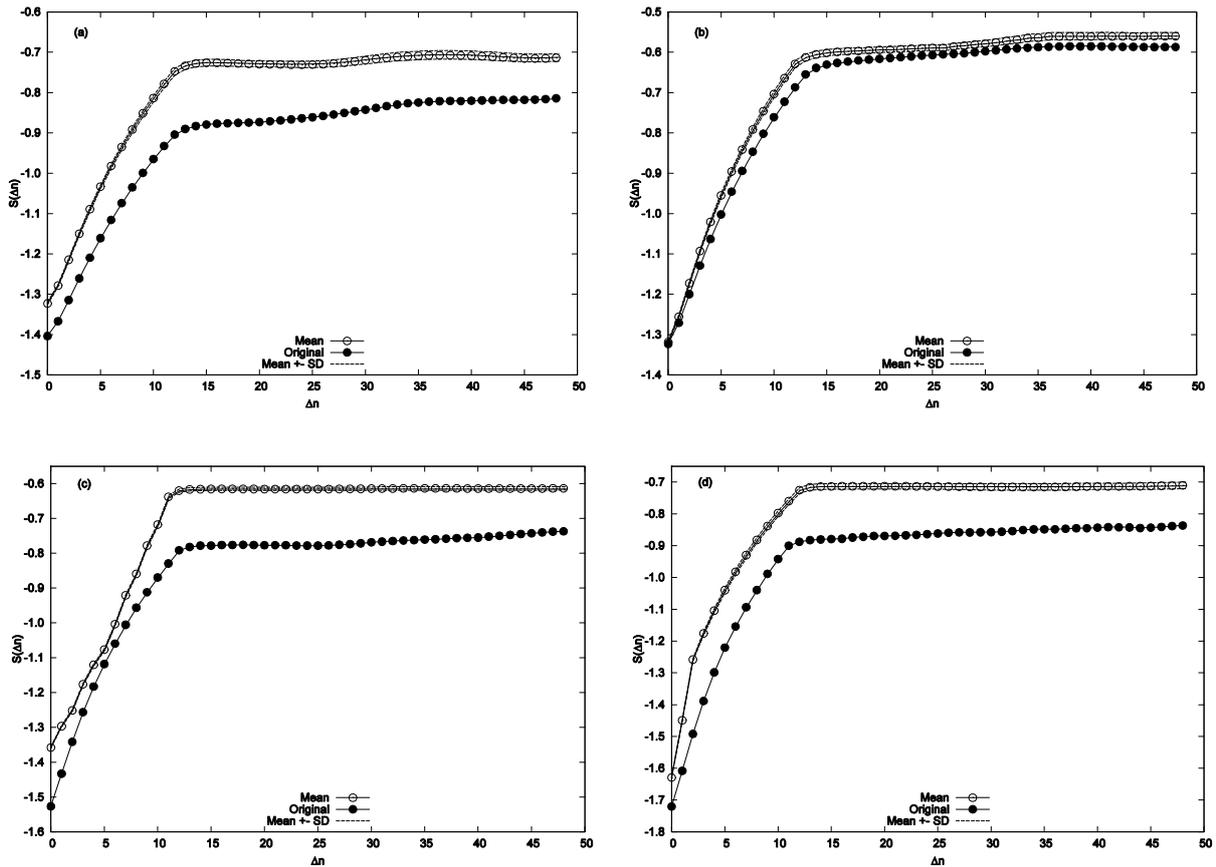

Figure 13: The mean value of $S(\Delta)$ of the surrogates with standard deviation. (a) Bhuj (b) Ahemadabad (c) Jabalpur (d) Birsa Munda Airport.

where $\mu$ and $\sigma$ are the mean and standard deviation of the characteristic computed from the surrogates and $\mu_{org}$ is the mean of the characteristic of the original data. It is estimated that we may reject the null hypothesis with 95% confidence level if $S > 2$. In other words if $S > 2$ the observed time series is not a realisation of a linear Gaussian stochastic process (Pavlos et al., 1999).

In Figure. 11 plotted the mean values of the fraction of false nearest neighbours of all surrogates and values one standard deviation away from the mean along that of the original data. The difference between the original data and the surrogates is evident in these figures. The significance of difference $S$ of the data is plotted in Figure. 12. It can be noted that the values of $S$ are remarkably higher than 2 for almost the entire range of values of $m$ considered. Therefore, we can safely reject the null hypothesis.

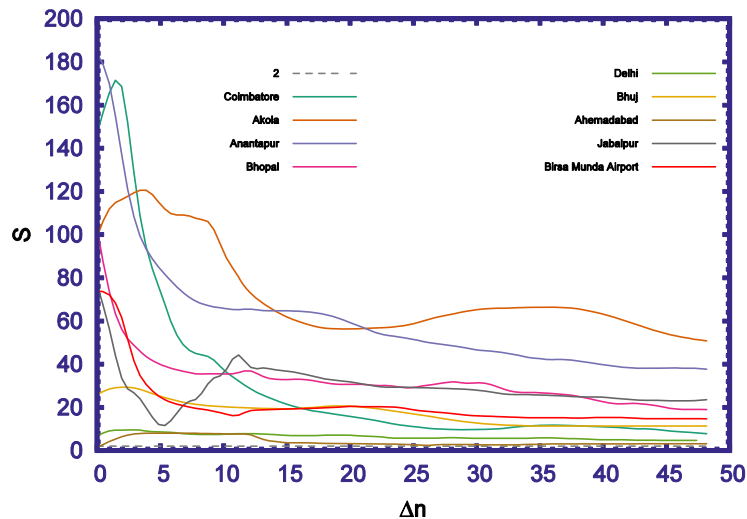

Figure 14: Plot of significance of difference $S$ versus $\Delta n$ for different locations

Next, we compare the original data with its surrogates with $S(\Delta n)$ of Eq.9 as the test statistic. Figure. 13 shows the curves of $S(\Delta n)$, of the surrogates with those of the original data with delay $\tau = 1$ Theiler window $\omega = 25$ and embedding dimension $m = 14$. A strong difference between the values of $S(\Delta n)$ corresponding to the original data and the surrogates is evident in all the locations. The significance of difference $S$ computed for all locations as shown in Figure. 14 is much above 2 for all $\Delta_n \leq 45$ and hence we can reject the null hypothesis with 95% confidence level.

## 6. Conclusion

In this work, we have carried out a detailed analysis of the daily mean wind speeds measured at various locations across Indian subcontinent with the objective to study deterministic content of the fluctuations. The analysis reveals that the underlying dynamics of wind speed oscillations is deterministic, low-dimensional and chaotic. The computed values of the fraction of false nearest neighbours and correlation dimension strongly suggest that the system is low dimensional independent of the location. The positive values of the maximum Lyapunov exponent across all locations show that the system is chaotic. The deterministic character was further established using surrogate data test using both structural and dynamics parameters. The results of this study shows that deterministic, low-dimensional and chaotic character pertains to all selected locations across Indian subcontinent.